\documentclass[12pt, epsfig, a4paper]{article}
\usepackage[dvips]{graphicx}
\usepackage{wrapfig}
\author{V.P. Berezovoj,
 G.I. Ivashkevych}
\title{Classical stochastic dynamics and extended $N = 4$ supersymmetric quantum mechanics.}
\begin{document}
\maketitle
\begin{center}
{\footnotesize National Scientific Center "Kharkov
Institute of Physics and
Technology"\\
Akademicheskaya str., 1, 61108, Kharkov, Ukraine\\ E-mail:
berezovoj@kipt.kharkov.ua}
\end{center}

{\footnotesize This work is aimed at demonstrating the possibility to construct new exactly-solvable stochastic systems by use of the extended supersymmetric quantum mechanics ($N=4~SUSY~QM$) formalism. A feature of the proposed approach consists in $N=4~SUSY~QM$ the fact that probability densities and so obtained new potentials, which enter the Langevin equation, have a parametric freedom. The latter allows one to change the potentials form without changing the temporal behavior of the probability density.}

\section{$N=4~SUSY~QM$}
Let us get started with a brief discussion of the $N=4~SUSY~QM$ structure and constructing within its framework isospectral Hamiltonians~\cite{berezovoj:berezovoj1,berezovoj:berezovoj2}. The $N=4~SUSY~QM$ Hamiltonian has the following form:

\begin{equation}
H_{\sigma _1 ,\sigma _2 }  = \frac{1}{2}(p^2  + V_2^2 (x) + \sigma _3^{(1)} V_2^1 (x)) \equiv \frac{1}{2}(p^2  + V_1^2 (x) + \sigma _3^{(2)} V_1^1 (x))
\end{equation}
where we have introduced ($\hbar = m = 1$):
\begin{equation}
V_i (x) = W'(x) + \frac{1}{2}\sigma _3^{(i)} \frac{{W''(x)}}{{W'(x)}}
\end{equation}
$W(x)$ is a superpotential, $\sigma _3^{(i)}$ are matrices which commute to each other and have the eigenvalues $\pm 1$
\begin{equation}
\sigma _3^{(1)}  = \sigma _3  \otimes 1,\sigma _3^{(2)}  = 1 \otimes \sigma _3
\end{equation}
Supercharges $Q_i$ of the extended supersymmetric mechanics form the algebra:
\begin{equation}
 \left\{ {Q_i ,\bar Q_k } \right\} = 2\delta _{ik} H; \left\{ {Q_i ,Q_k } \right\} = 0; i,k = 1,2
\end{equation}
And admit the form:
\begin{equation}
\begin{array}{l}
 Q_i  = \sigma _ - ^{(i)} (p + iV^{(i + 1)} (x)),~
 \bar Q_i  = \sigma _ + ^{(i)} (p - iV^{(i + 1)} (x))
 \end{array}
\end{equation}

Constructing isospectral Hamiltonians within the $N=4~SUSY~QM$ is based on the fact that four Hamiltonians cast into one supermultiplet. Let us take as an initial one of the Hamiltonians
\begin{equation}
\begin{array}{l}
 H_{\sigma _1 ,\sigma _2 }  = \frac{1}{2}(p - i\sigma _1 V_{\sigma _2 } (x))(p + i\sigma _1 V_{\sigma _2 } (x)) + \varepsilon  \equiv
 \frac{1}{2}(p - i\sigma _2 V_{\sigma _1 } (x))(p + i\sigma _2 V_{\sigma _1 } (x)) + \varepsilon  \\
 \end{array}
\end{equation}
where
\begin{equation}
V_{\sigma _2 } (x) = W'(x) + \frac{1}{2}\sigma _2 {{W''(x)} \mathord{\left/
 {\vphantom {{W''(x)} {2W'(x)}}} \right.
 \kern-\nulldelimiterspace} {W'(x)}},
\end{equation}
and $\varepsilon$ is the so-called factorization energy. In what follows the energy level will be counted from $\varepsilon$.

Consider an auxiliary equation:
\begin{equation}
H_{\sigma _1 }^{\sigma _2 } \varphi (x) = \varepsilon \varphi (x)
\end{equation}
Its general solution $\varphi (x,\varepsilon ,c)$ is the linear combination of two independent solutions $\varphi ^{(1)} (x,\varepsilon )$ and $\varphi ^{(2)} (x,\varepsilon )$. The expression for $W(x)$ in terms of $\varphi (x,\varepsilon ,c)$ has the from of:
\begin{equation}
W(x) = \frac{{\sigma _1 }}{2}\ln \left(1 + \lambda \int\limits_{x_i }^x {dx'\left[ {\varphi (x',\varepsilon ,c)} \right]^{2\sigma _1 \sigma _2 } } \right)
\end{equation}
where $\lambda ,x_i$ are two new parameters.

For definiteness we will consider $\sigma _1  = \sigma _2  =  - 1$. Then, relations in the above can be presented in the form of:
\begin{equation}
 H_ + ^ -   = H_ - ^ -   - \frac{{d^2 }}{{dx^2 }}\ln \varphi (x,\varepsilon ,c),~ \psi _ + ^ -  (x,E) = \frac{1}{{\sqrt {2(E - \varepsilon )} }}\frac{{W\left\{ {\varphi (x,\varepsilon ,c)\psi _ - ^ -  (x,E)} \right\}}}{{\varphi (x,\varepsilon ,c)}}
\end{equation}
\begin{equation}
\begin{array}{c}\label{berezovoj:2}
 H_ + ^ +   = H_ - ^ -   - \frac{{d^2 }}{{dx^2 }}\ln (1 + \lambda \int\limits_{x_i }^x {dx'\left[ {\varphi (x',\varepsilon ,c)} \right]^2 } ) \\
 \psi _ + ^ +  (x,E) = \psi _ - ^ -  (x,E) - \frac{{\lambda \varphi (x,\varepsilon ,c)}}{{1 + \lambda \int\limits_{x_i }^x {dx'\left[ {\varphi (x',\varepsilon ,c)} \right]^2 } }}\int\limits_{x_i }^x {dx'^2 } \varphi (x',\varepsilon ,c)\psi _ - ^ -  (x',E) \\
 \end{array}
\end{equation}

In the case when $\varphi (x,\varepsilon ,c)$ is a ground state eigenfunction of the initial Hamiltonian, wave function of ground state $H_ + ^ + $
has a form:
\begin{equation}\label{berezovoj:wave}
\psi _ + ^ +  (x,E_0 ) = N_0 \frac{{\varphi (x,\varepsilon )}}{{1 + \lambda \int\limits_{x_i }^x {dx'\left[ {\varphi (x',\varepsilon )} \right]^2 } }}
\end{equation}

\section{ Construction of stochastic models associated with $N=4~SUSY~QM$.}

The Fokker-Planck equation is equivalent to the Langevin equation,
however the Fokker-Planck is used more widely in physics, since it
is formulated in more common for the probability density $m_t^ \pm
(x,x_0 )$ language. The Fokker-Planck equation takes the
form~\cite{berezovoj:risken,berezovoj:junker}:
\begin{equation}
\begin{array}{c}
 \frac{\partial }{{\partial t}}m_t^ \pm  (x,x_0 ) = \frac{D}{2}\frac{{\partial ^2 }}{{\partial x^2 }}m_t^ \pm  (x,x_0 ) \mp \frac{\partial }{{\partial x}}\Phi (x)m_t^ \pm  (x,x_0 ) \\
 m_{t = 0}^ \pm  (x,x_0 ) = \left\langle {\delta (x - x_0 )} \right\rangle ;~ U_ \pm  (x) =  \mp \int\limits_0^x {dz}  \Phi (z) \\
 \end{array}
\end{equation}
with $U_ \pm  (x)$ to be the potential entering the Langevin
equation. The Fokker-Planck equation describes the stochastic
dynamics of particles in potentials $U_ + $ and $U_ -   =  - U_ + $.
Substituting
\begin{equation}
m_t^ \pm  (x,x_0 ) = \exp \left\{ { - \left[ {U_ \pm  (x) - U_ \pm  (x_0 )} \right]/D} \right\}K_ \pm  (x,t)
\end{equation}
the Fokker-Planck equation transforms into the Schrodinger equation
with imagine time:
\begin{equation}
\begin{array}{c}
  - D\frac{\partial }{{\partial t}}K_ \pm  (x,t) = \left\{ { - \frac{{D^2 }}{2}\frac{{\partial ^2 }}{{\partial x^2 }} + \frac{1}{2}\left[ {\Phi ^2 (x) \pm  D\Phi '(x)} \right]} \right\}K_ \pm  (x,t) = H_{\pm}K_{\pm}(x,t)\\
 K_ \pm  (x,t) = \left\langle {x\left| {\exp \left\{ { - tH_\pm/D} \right\}} \right|x_0 } \right\rangle
 \end{array}
\end{equation}
in which the diffusion constant can be treated as the ''Planck''
constant. As in the previous case, let us consider $H_ - ^ -  $ to
the initial Hamiltonian.
\begin{equation}
\begin{array}{c}
 H_-^-= \frac{1}{2}\left[p^2+\left(V_{2}^{(-)}(x)\right){}^{2}-DV_2^{(-)}{}'(x) \right] \\
 V_2^{( - )} (x) = W'(x) - \frac{D}{2}\frac{d}{{dx}}\ln \left| {W'(x)} \right| \equiv  - D\frac{d}{{dx}}\ln \varphi (x,\varepsilon ,c) \\
 \end{array}
\end{equation}

The force entering the corresponding Langevin equation:
\begin{equation}
\begin{array}{c}
 F(x) =  - \frac{{dU_ - ^ -  (x)}}{{dx}} =  D\frac{d}{{dx}}\ln \varphi (x,\varepsilon ,c),~ U_ - ^ -  (x) =  - D\ln \varphi (x,\varepsilon ,c) \\
 \end{array}
\end{equation}
Then, in the same to the $N=2~SUSY~QM$ way, there is the relation $U_ + ^ -   =  - U_ - ^ -$. It directly follows from the fact that $H_ + ^ -$ is obtained from $H_ - ^ -$ with changing $V_2^{( - )} (x)$ to $-V_2^{( - )} (x)$.

Hence, there is the relation between the stochastic dynamics in a potential and that of in an inverse potential.

Further consideration is based on an important property of $N=4~SUSY~QM$, which consists in having the symmetry of $H_{\sigma _1 }^{\sigma _2 }$ under $
\sigma _1  \leftrightarrow \sigma _2$. It leads to:
\begin{equation}
H_ + ^ -  (x,p) = \frac{1}{2}\bar Q_1^{( - )} Q_1^{( - )}  \equiv \frac{1}{2}Q_2^{( + )} \bar Q_2^{( + )}  = H_ - ^ +  (x,p)
\end{equation}
If the first equation points to the expression of $H^{( + )}$ in terms of $Q_1 (\bar Q_1 )$, the second one is $H^{( - )}$ in terms of supercharges $
Q_2 (\bar Q_2 )$. Though the supercharges of $H^{( + )}$ and $H^{( - )}$ are essentially different:
\begin{equation}
\begin{array}{c}\label{berezovoj:1}
 H_ + ^ -  (x,p) = \frac{1}{2}Q_2^{( + )} \bar Q_2^{( + )}  = \frac{1}{2}\left[ {p^2  + \left( {V_1^{( + )} (x)} \right){}^2  - DV_1^{( + )}{}^\prime  (x)} \right] \\
 V_1^{( + )} (x) = \frac{d}{{dx}}(W + \frac{D}{2}\ln \left| {W'(x)} \right|) \equiv D\frac{d}{{dx}}\ln \left| {\frac{{\varphi (x,\varepsilon ,c)}}{{1 + \lambda \int\limits_{x_i }^x {dx'\left[ {\varphi (x',\varepsilon ,c)} \right]^2 } }}} \right|=\frac{dU_-^+}{dx} \\
 \end{array}
\end{equation}

From equations (\ref{berezovoj:1}) follows that at the quantum level the Hamiltonians $H_ - ^ + $ and $H_ + ^ -$ possesses the same spectrum and wave functions, thus the corresponding to them $K(x,t)$ are the same. At the same time the corresponding to these Hamiltonians stochastic models are described by essentially different potentials, so their $m_t (x,x_0 )$ are different. Further, the potential $U_-^+(x)$ has a nontrivial parametric dependence on $\lambda $, that corresponds to having the family of stochastic models, the probability densities of which have the same temporal dependence. Having the parametric freedom allows one to change the potential form that is unexpected in the case. In particular, as it has been noticed in literature, physical quantities such as the time of passing a peak of potentials depend essentially on local changes of the potential barrier.
Looking at $H_ + ^ +$, we note that it follows from $H_ - ^ +$ with changing $
V_1^{( + )} (x)$ to  $- V_1^{( + )} (x)$, as it also takes place in $N=2$ supersymmetry. In its turn the potential of this stochastic system is $U_ + ^ +  (x) =  - U_ - ^ +  (x)$. The wave functions which are used to calculate $K_ + ^ +  (x,t)$ have the form of (\ref{berezovoj:2}) in the case.

Getting back to the parametric dependence of $m_{ - ,t}^ +  (x,x_0
)$ and $ m_{ + ,t}^ +  (x,x_0 )$, it should be noted that the
normalization condition could lead to fixing the $\lambda$. The same
does not happen in the case of $m_{ + ,t}^ +  (x,x_0)$, if instead
of  $\varphi (x,\varepsilon ,c)$ we will use the normed wave
function (say, the ground state wave function) of the initial
Hamiltonian $H_ - ^ -  $. From definition:
\begin{equation}
\begin{array}{c}
m_{ + ,t}^ +  (x,x_0 ) = \frac{{\varphi (x,\varepsilon ,c)}}{{1 +
\lambda \int\limits_{ - \infty }^x {dx'\left[ {\varphi
(x',\varepsilon ,c)} \right]^2 } }} \times\\
\times \frac{{1 + \lambda \int\limits_{ - \infty }^{x_0 } {dx'\left[
{\varphi (x',\varepsilon ,c)} \right]^2 } }}{{\varphi (x,\varepsilon
,c)}}\sum\limits_{n = 0} {e^{ - \frac{{E_n }}{D}t} \psi _ + ^ +
(x,E_n )\psi _ + ^ +  (x_0 ,E_n )}
\end{array}
\end{equation}
in view of eq. (\ref{berezovoj:wave}) it follows  that
\begin{equation}
\begin{array}{c}
\int\limits_{ - \infty }^\infty  {dxm_{ + ,t}^ +  (x,x_0 )} =
\int\limits_{ - \infty }^\infty  {dxm_{ + ,t = 0}^ +  (x,x_0 )} =\\
=(\lambda  + 1)\int\limits_{ - \infty }^\infty  {dx} \left[
\frac{{\varphi (x,\varepsilon ,c)}}{{1 + \lambda \int\limits_{ -
\infty }^x {dx'\left[ {\varphi (x',\varepsilon ,c)} \right]^2 } }}
\right]^2= 1
\end{array}
\end{equation}
It means that the unique restriction to $\lambda$ is $ \lambda  \ne
- 1$, that as it was early noticed corresponds to the absence of
singularities at the quantum-mechanical level. It is hard to make an
analogous study for $m_{ - ,t}^ +  (x,x_0 )$ in a general case.
However, for the Ornstein-Uhlenbeck process the normalization
condition does not remove the parametric ambiguity at the definite
choice of $x_0$.

\section{The Ornstein-Uhlenbeck process}

Let us demonstrate the proposed scheme of getting new stochastic
models with the well-known Ornstein-Uhlenbeck process. The
Fokker-Planck equation which describes the Ornstein-Uhlenbeck leads
to a quantum-mechanical potential with the harmonic oscillator
potential. The factorization energy coincides with the ground state
energy in the case with the harmonic oscillator potential.
\begin{equation}
\left( {\frac{{p^2 }}{2} + \frac{{\omega ^2 }}{2}x^2 } \right)\varphi (x,\varepsilon ) = \varepsilon \varphi (x,\varepsilon )
\end{equation}
\begin{equation}
E_n  = nD\omega ,  \psi _-^- (x,E_n) = \left(\omega/D \right)^{{\raise0.7ex\hbox{$1$} \!\mathord{\left/
 {\vphantom {1 2}}\right.\kern-\nulldelimiterspace}
\!\lower0.7ex\hbox{$2$}}} H_n (x\sqrt {\omega/D} )e^{ - \frac{\omega }{D}x^2 } ,   n = 0,1,...
\end{equation}
where $\varepsilon  = \frac{{D\omega }}{2}$ (the energy counts from $\varepsilon$). The potential entering the Langevin equation:
\begin{equation}U_ - ^ -  (x) = \omega x^2/2 = D\xi ^2/2,~\xi  = \sqrt {\omega/D} x.
\end{equation}

For the Ornstein-Uhlenbeck process it is easy to see that, at least for the case of $x_0  = 0$, the normalized condition does also not fix the value of $\lambda$. For an arbitrary stochastic process the same is hard to prove.
Calculating $m_{ + t}^ +  (\xi ,\xi _0 )$ we use the wave functions (\ref{berezovoj:2}) . First of all it has to be pointed out of having the equilibrium value of $m_{ + ,t \to \infty }^ +  (\xi ,\xi _0 )$:
\begin{equation}
m_{ + ,t \to \infty }^ +  (\xi ,\xi _0 ) = (\lambda  + 1)\left( {\frac{\omega }{{\pi D}}} \right)^{{\raise0.7ex\hbox{$1$} \!\mathord{\left/
 {\vphantom {1 2}}\right.\kern-\nulldelimiterspace}
\!\lower0.7ex\hbox{$2$}}} \frac{{e^{ - \xi ^2 } }}{{(1 + {\lambda  \mathord{\left/
 {\vphantom {\lambda  2}} \right.
 \kern-\nulldelimiterspace} 2} + {\lambda  \mathord{\left/
 {\vphantom {\lambda  2}} \right.
 \kern-\nulldelimiterspace} 2}\Phi (\xi ))^2 }},~\Phi (\xi ) = \frac{2}{{\sqrt \pi  }}\int\limits_0^\xi  {dt e^{ - t^2 } }
\end{equation}

As it was mentioned in the above, the normalized condition of the probability density does not remove the $\lambda$ - parametric freedom. Using the wave functions $\Psi _ + ^ +  (x,E_n )$, which are expressed in terms of $\Psi _ - ^ -  (x,E_n )$, after simple but tedious calculations we get the expression for the probability density:
\begin{equation}
 \begin{array}{l}
 m_{ + t}^ +  (\xi ,\xi _0 ) = \lambda \left( {\frac{\omega }{{\pi D}}} \right)^{{\raise0.7ex\hbox{$1$} \!\mathord{\left/
 {\vphantom {1 2}}\right.\kern-\nulldelimiterspace}
\!\lower0.7ex\hbox{$2$}}} \frac{{e^{ - \xi ^2 } }}{{(1 + {\lambda  \mathord{\left/
 {\vphantom {\lambda  2}} \right.
 \kern-\nulldelimiterspace} 2} + {\lambda  \mathord{\left/
 {\vphantom {\lambda  2}} \right.
 \kern-\nulldelimiterspace} 2}\Phi (\xi ))^2 }} + \frac{1}{2}\left( {\frac{\omega }{D}} \right)^{{\raise0.7ex\hbox{$1$} \!\mathord{\left/
 {\vphantom {1 2}}\right.\kern-\nulldelimiterspace}
\!\lower0.7ex\hbox{$2$}}} \frac{d}{{d\xi }}\frac{1}{{1 + {\lambda  \mathord{\left/
 {\vphantom {\lambda  2}} \right.
 \kern-\nulldelimiterspace} 2} + {\lambda  \mathord{\left/
 {\vphantom {\lambda  2}} \right.
 \kern-\nulldelimiterspace} 2}\Phi (\xi )}} \times  \\
 \left[ {\left( {1 + {\lambda  \mathord{\left/
 {\vphantom {\lambda  2}} \right.
 \kern-\nulldelimiterspace} 2} + {\lambda  \mathord{\left/
 {\vphantom {\lambda  2}} \right.
 \kern-\nulldelimiterspace} 2}\Phi (\kappa )} \right)\left( {1 + \Phi \left( {\frac{{\xi  - \kappa z}}{{\sqrt {1 - z^2 } }}} \right)} \right) - \frac{\lambda }{{\sqrt \pi  }}\int\limits_{ - \infty }^\xi  {d\xi 'e^{ - \xi '^2 } \left[ {1 + \Phi \left( {\frac{{\kappa  - \xi 'z}}{{\sqrt {1 - z^2 } }}} \right)} \right]} } \right] \\
 \end{array}
\end{equation}

The form of $m_{ + ,t}^ +  (\zeta ,\zeta _0 )$ can be viewed from the following plots under the following parametrization $\zeta _0  = 0$,
 $z = e^{ - \omega t}$.

\begin{figure}
\begin{minipage}{.45\textwidth}
\centering
\includegraphics[bb= 0 0 600 600, width=.9\textwidth]{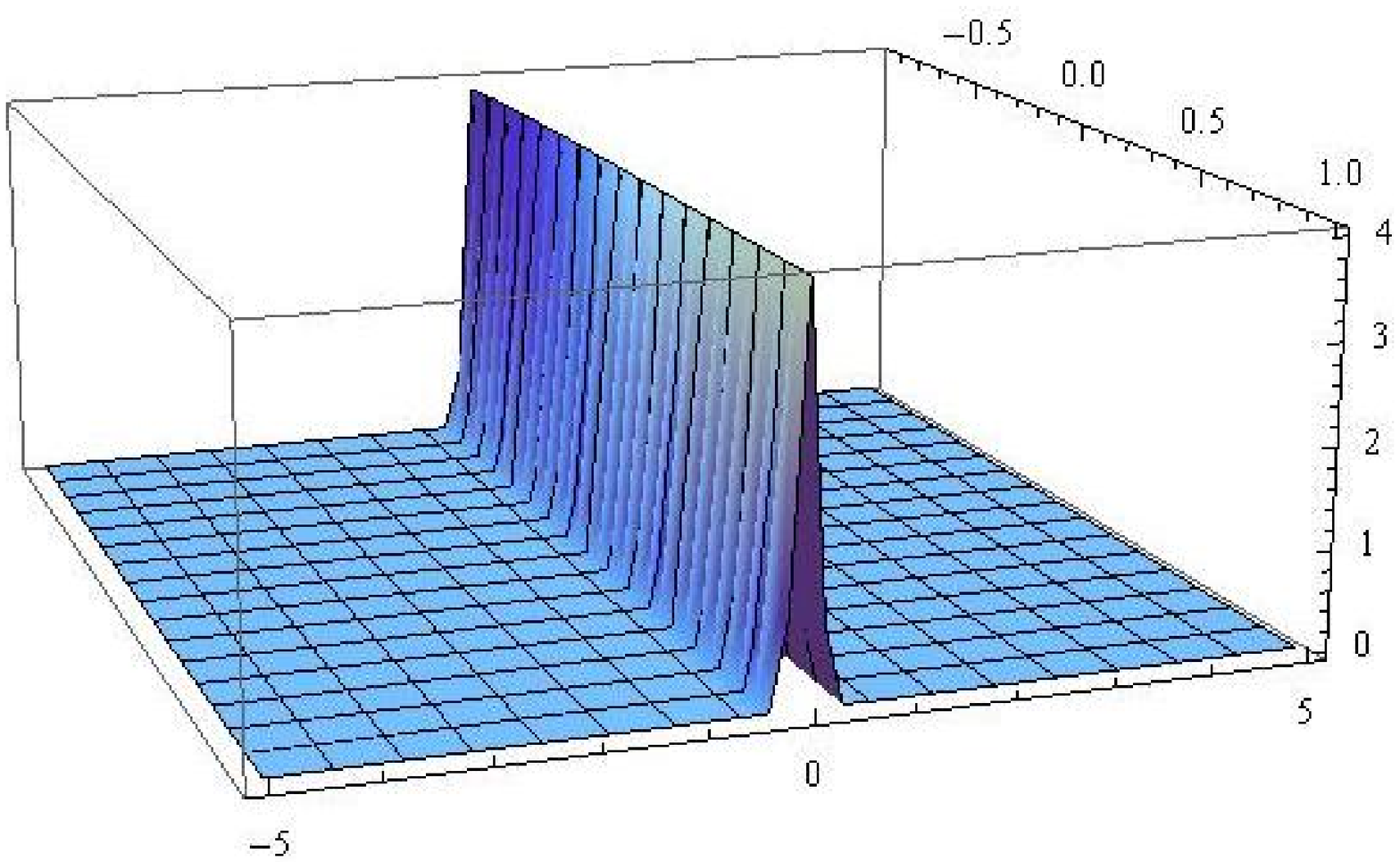}
\caption{Dependence of $m_{ + ,t}^ +  (x,0)$ on $x$ and $\lambda $  at $z=1$.} \label{berezovoj:fig:1}
\end{minipage}
 \rule{.05\textwidth}{0pt}
\begin{minipage}{.45\textwidth}
\centering
\includegraphics[bb= 0 0 600 600 ,width=.9\textwidth]{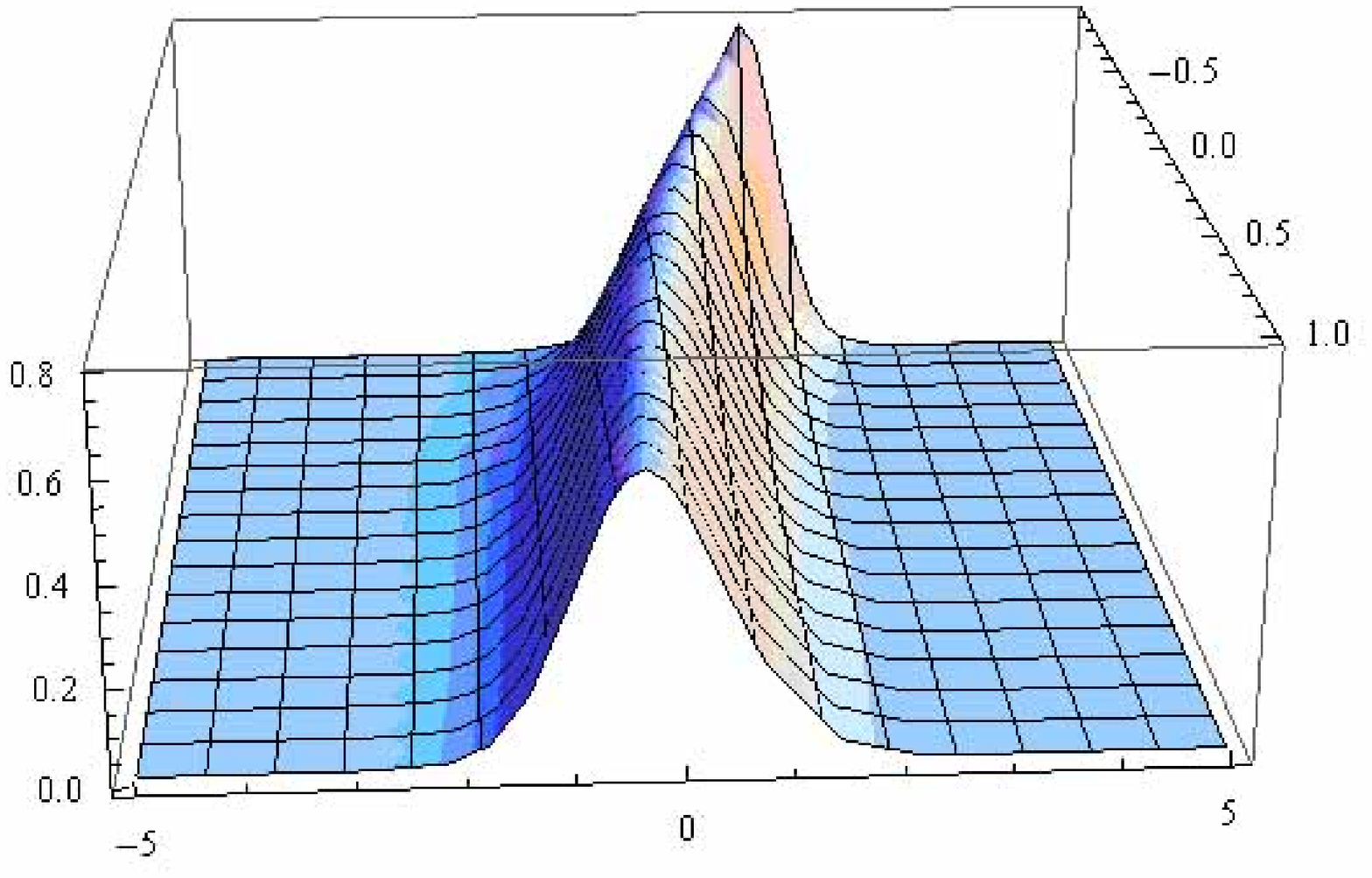}
\caption{Dependence of $m_{ + ,t}^ +  (x,0)$ on $x$ and $\lambda $  at $z=0$.} \label{berezovoj:fig:2}
\end{minipage}
\end{figure}

\begin{figure}
\begin{minipage}{.45\textwidth}
\centering
\includegraphics[bb= 0 0 600 600, width=.9\textwidth]{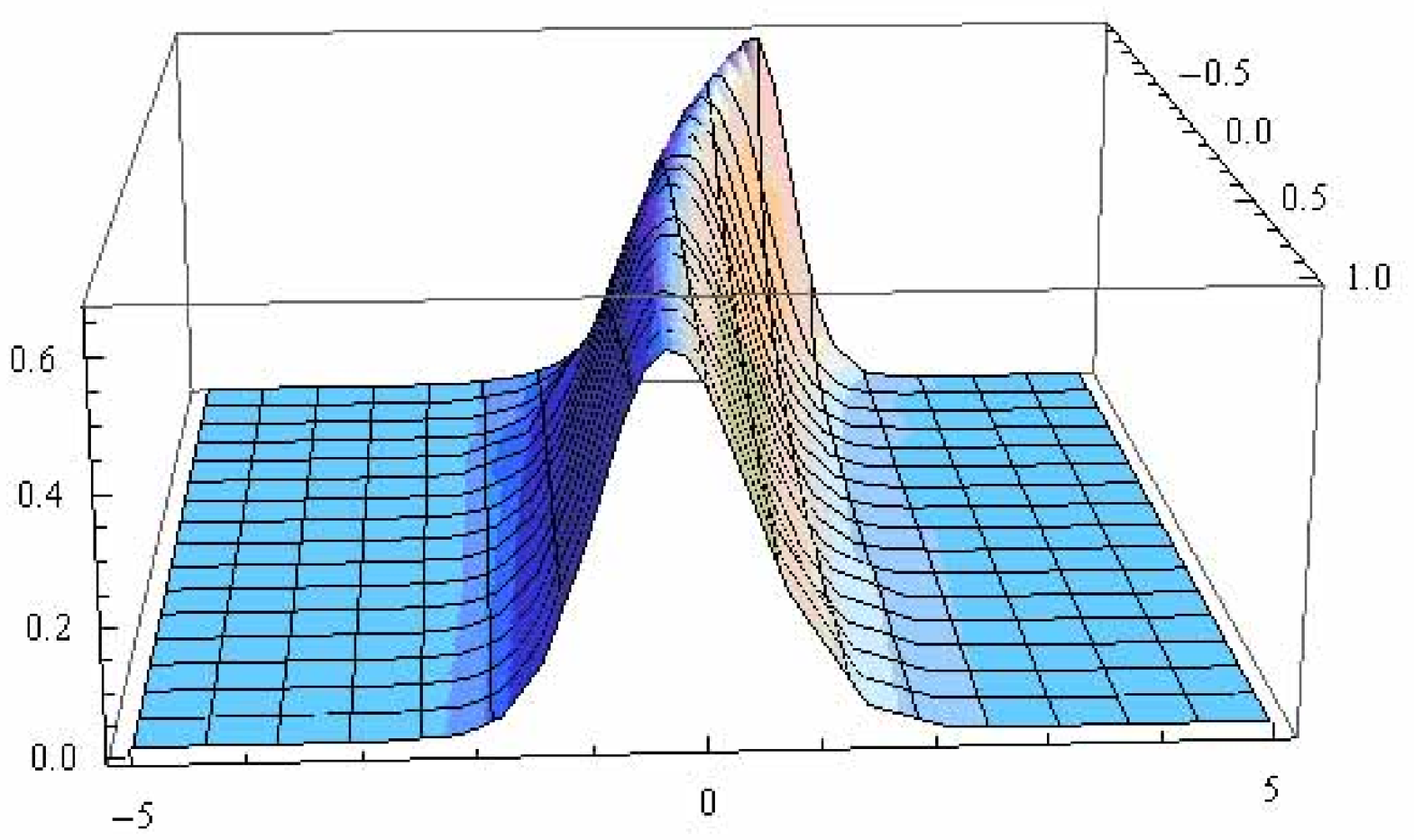}
\caption{Dependence of $m_{ + ,t}^ +  (x,0)$ on $x$ and $\lambda $
at $z=0.25$.} \label{berezovoj:fig:3}
\end{minipage}
 \rule{.05\textwidth}{0pt}
\begin{minipage}{.45\textwidth}
\centering
\includegraphics[bb= 0 0 600 600 ,width=.9\textwidth]{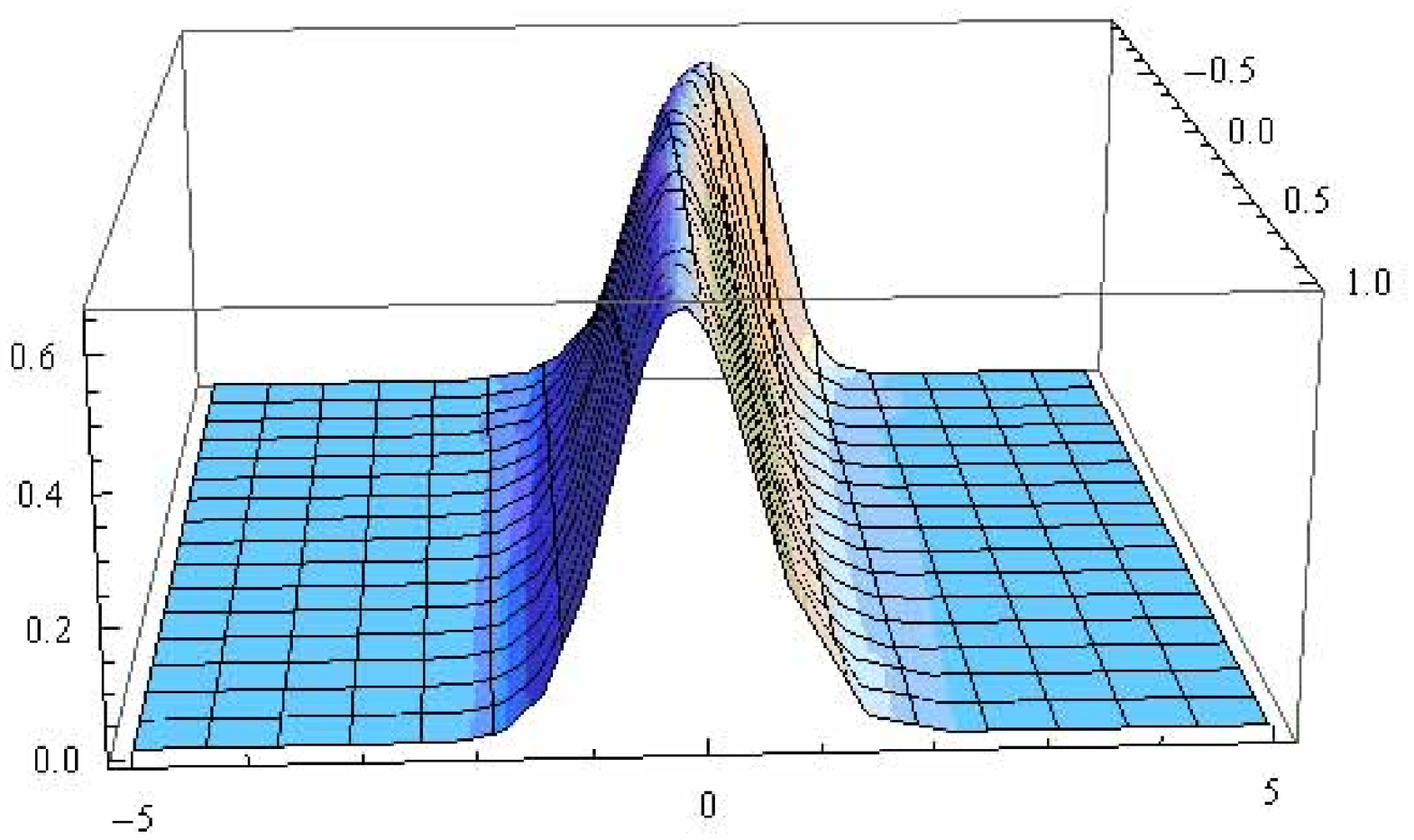}
\caption{Dependence of $m_{ + ,t}^ +  (x,0)$ on $x$ and $\lambda $
at $z=0.5$.} \label{berezovoj:fig:4}
\end{minipage}
\end{figure}

Fig.(\ref{berezovoj:fig:1}-\ref{berezovoj:fig:4}) show the results
of numerical evaluation of the dependence of $m_{ + ,t}^ +  (x,0)$
on $x$ and $\lambda $  at different $z$. Thus, when $z=1$ $m_{ +
,t}^ +  (x,0)$ has well pronounced $\delta$-like shape in
concordance with initial conditions. With decreasing of $z$  $m_{ +
,t}^ +  (x,0)$ changes its shape and this changes are especially
significant at $\lambda \rightarrow -1$. In this case the shift of
the maximum of probability density occurs, as well as emergence of
substantial asymmetry.


\section{Conclusions.}

The described procedure of the obtaining of exactly-solved
stochastic models allows to use the results of numerous works
concerning exactly-solved as well as quasi-exactly-solved quantum
mechanical problems. The distinctive feature of this approach is an
existence of parametric freedom in potentials, that enter the in the
Langevin equation, as well as in transitional probability densities.
This situation takes place at any modification in the spectrum of
initial Hamiltonian. It's important to mention, that normalization
condition for $m_{ - ,t}^ +  (x,x_0 )$, most probably, could be
fulfilled only when certain relation between $x_0 $ and $\lambda $
exists. This means that stochastic model with $U_ - ^ +  (x)$ is
''bad''. At the same time $m_{ + ,t}^ +  (x,x_0 )$ is a ''good''
transitional probability density, which preserve the parametric
freedom.

As is well known, \cite{berezovoj:berej,berezovoj:malah}, local
modifications of the shape of the potential in Langevin equation
could lead to substantial changes of such quantities as times  of
the passing through barrier and time of live in metastable state.
Although this quantities are mainly determined by first non-zero
energy level in spectrum of $H_{\sigma _1 }^{\sigma _2 }$, the
$\lambda$-freedom allows for significant variations of their values.
This especially important for potentials with several local minima,
which could emerge when constructing isospectral Hamiltonians with
factorization energy  $\varepsilon  < E_0 $. In this case potentials
$U_{\sigma _1 }^{\sigma _2 } $ with two wells emerge, symmetric, as
well as asymmetric and $\lambda$-freedom reveals in substantial
modification of the shape and height of the barriers, what leads to
changes in the rate of the inter-well transitions.

Authors are proud to express their thanks to Yu.L.Bolotin and
A.V.Olemskoi for attention to this work and constructive
discussions. Work was supported by INTAS (2006-7928) and NASU-RFFR
$\#$38/50-2008 grants.

\end{document}